\documentclass[conference]{IEEEtran}
\IEEEoverridecommandlockouts
\usepackage{cite}
\usepackage{subcaption}
\usepackage{amsmath,amssymb,amsfonts}
\usepackage{algorithm}
\usepackage{algpseudocode}
\usepackage{booktabs}
\usepackage{multirow}
\usepackage{graphicx}
\usepackage{textcomp}
\usepackage{xcolor}
\usepackage{url}
\def\BibTeX{{\rm B\kern-.05em{\sc i\kern-.025em b}\kern-.08em
    T\kern-.1667em\lower.7ex\hbox{E}\kern-.125emX}}

\usepackage{xspace}
\usepackage{marginnote}

\begin{document}

\title{Proactive Load-Shaping Strategies with Privacy-Cost Trade-offs in Residential Households based on Deep Reinforcement Learning
\thanks{RZ is funded by China Scholarship Council - The University of Manchester (CSC-UoM) joint Scholarship Programme. This work was also partially supported by EPSRC, under the project EnnCore: End-to-End Conceptual Guarding of Neural Architectures (EP/T026995).}
}
\author{\IEEEauthorblockN{Ruichang Zhang\IEEEauthorrefmark{1}, Youcheng Sun\IEEEauthorrefmark{1}, and
Mustafa A. Mustafa\IEEEauthorrefmark{1}\IEEEauthorrefmark{2}}
\IEEEauthorblockA{\IEEEauthorrefmark{1}Department of Computer Science, The University of Manchester, Oxford Road, Manchester, M13 9PL, UK\\
\IEEEauthorrefmark{2}COSIC, KU Leuven, Leuven, 3001, Belgium}
Email: ruichang.zhang-2@postgrad.manchester.ac.uk, youcheng.sun@manchester.ac.uk, mustafa.mustafa@manchester.ac.uk}

\maketitle

\begin{abstract}
Smart meters play a crucial role in enhancing energy management and efficiency, but they raise significant privacy concerns by potentially revealing detailed user behaviors through energy consumption patterns. Recent scholarly efforts have focused on developing battery-aided load-shaping techniques to protect user privacy while balancing costs. This paper proposes a novel deep reinforcement learning-based load-shaping algorithm (PLS-DQN) designed to protect user privacy by proactively creating artificial load signatures that mislead potential attackers. We evaluate our proposed algorithm against a non-intrusive load monitoring (NILM) adversary. The results demonstrate that our approach not only effectively conceals real energy usage patterns but also outperforms state-of-the-art methods in enhancing user privacy while maintaining cost efficiency. 
PLS-DQN reduces the \(F_1\) score for the NILM adversary's classification results by 95\% and 92\% for the on/off status of two common appliances: kettle and toaster, respectively. When compared to the state-of-the-art DDQL-MI model, PLS-DQN not only lowers the \(F_1\) score by 84\% and 79\% respectively but also achieves a 42\% reduction in household electricity costs.
\end{abstract}


\section{Introduction}
Smart meters (SM) are pivotal components in the evolution of modern smart homes. These devices, which provide real-time data on electricity consumption, have been recognized for their potential to enable dynamic pricing and enhanced smart grid management\cite{efthymiou2010smart,sankar2012smart}. SM systems act as the cornerstone of the transformation, providing real-time energy consumption data, and enabling dynamic pricing models that can fluctuate based on demand and supply\cite{mcdaniel2009security}. SMs can monitor customer power usage information in real-time and upload it to the smart grid. However, the data they collect can be so detailed that it could reveals specific user behaviors as highlighted in~\cite{greveler2012multimedia,asghar2017smart,pan2019you,chen2023control, XJ2023TCN}. It can reveal the daily activities and habits of household occupants, potentially leading to privacy leakage.

Each electrical appliance has a unique power signature that can be identified. Non-intrusive load monitoring (NILM), also known as energy disaggregation, is a computational technique that disaggregates the total energy consumption of a building or household into individual appliances usage~\cite{kelly2015neural, eskandarnia2022deep}. NILM algorithms have the potential to infer sensitive personal information by analyzing the usage patterns of certain appliances.
While measures to protect this data can be implemented, they often increase energy costs~\cite{cuijpers2013smart}. This balance between privacy and cost forms the research problem this work tackles: \textit{How can SM systems be both private and cost-effective?}

Although existing works \cite{gomez2014smart,yang2012minimizing,varodayan2011smart,tan2013increasing,tan2017privacy,li2018information,shateri2020privacy,shateri2021privacy,li2023research} propose novel environmental settings and algorithms to solve the privacy-cost trade-off problem of SM data, there are still some shortcomings such as privacy metrics and environment settings used in practical applications. These limitations are summarized next.

\textit{Flatness privacy metrics:} Ideally, flattening the load could hide the signatures of appliances and patterns of household occupancy. However, full flatness during all times of a time cycle is hard to achieve due to the physical resource constraint. Only flatness by time period \cite{giaconi2020privacy} or soft constraint-based learning algorithm \cite{sun2017smart,sun2016ev} can be realized, which still has the risk for privacy leakage. 

\textit{Mutual information (MI) privacy metrics:} Utilizing MI as a privacy metric between the masked load and the original load could demonstrate strong statistical dependence for the entire sequence \cite{shateri2021privacy,li2018information}. However, even a low MI value does not assure statistical correlation among subsequent data. In the context of SM data, the operational duration of specific appliances occupies only a small fraction of the day during the recording period. Consequently, when faced with an adversary utilizing NILM techniques, MI-based algorithms cannot guarantee privacy for these brief consumption intervals. 

\textit{Environmental settings:} Existing literature \cite{li2018information,shateri2020privacy,shateri2021privacy,li2023research} neglect the necessity of keeping battery levels consistent before and after cycles. After one decision period, a change in the battery level affects the observation space of the privacy-preserving algorithm, leading to sub-optimal results in the next cycle. Particularly, when the battery is sourced from an electric vehicle, this environmental setting is crucial. It ensures that the vehicle is always ready for immediate use after the cycle, providing reliability and convenience to users. 

To address these limitations, our work makes the following contributions:
\begin{itemize}
\item 
We propose PLS-DQN -- a novel proactive load-shaping strategy based on Deep Q-Network (DQN) algorithm in deep reinforcement learning (DRL) that introduces artificial load patterns and masks original load patterns to enhance privacy in SM data. Our approach provides a robust method for maintaining user privacy against sophisticated data analysis techniques like NILM. To the best of our knowledge, this is the first study to integrate proactive load-shaping strategies with artificial load patterns specifically designed to mislead advanced NILM techniques used by potential attackers. 

\item 
We introduce a novel reward-shaping strategy in DRL to optimize the proactive load-shaping algorithms for SM data protection. It specifically rewards strategies that lead to the creation of artificial load patterns and masking of existing load signatures, encouraging the model to prioritize privacy-centric behaviors. 

\item 
We propose a battery consistency mechanism that strategically manages the discharge/recharge cycles within the load-shaping process, ensuring that the battery starts and ends with similar levels. It enhances the effectiveness of our load-shaping strategy for subsequent operating cycles. 
\end{itemize}
The rest of the paper is organized as follows: In Section~\ref{sec:model}, we model the environment with a Markov decision process (MDP). In Section~\ref{sec:methodology}, we describe PLS-DQN and the adversary NILM algorithm. The evaluation results are presented in Section~\ref{sec:results}. Finally, Section~\ref{sec:conclusion} concludes the paper.

\section{Problem Formulation}
\label{sec:model}
\subsection{System Model}

\begin{figure}[t]
\centering
\includegraphics[width=0.5\textwidth]{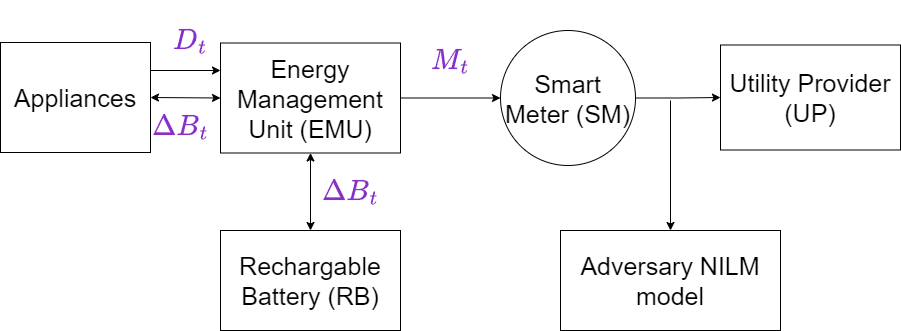}
\caption{System Model.}
\label{fig: system model}
\end{figure}

In this section, we set a time-discrete household SM system model (see Fig.~\ref{fig: system model}). A rechargeable battery (RB) serves as the electricity storage device, which could be used for hiding the load signature in energy consumption data. An energy management unit (EMU) is used to implement load-shaping strategies by adjusting the RB charging/discharging actions. 

Let \( D_t \) represent the user’s demand load, which reflects the total power required by various appliances at a specific time of the day \( t \). \( M_t \) denotes the masked load generated by EMU and reported to the utility provider (UP). The time index \( t \) belongs to the set \( {\mathcal{T} = \{1, \ldots, T\}} \). \( B_t \) represents the amount of electricity in RB at any time \( t \). \( \Delta B_t \) represents the energy exchange from the RB through EMU's control strategy. Thus, \( M_t \) and \( B_{t+1} \) is determined by \({M_t=D_t + \Delta B_t}\) and \({B_{t+1} = B_t + \Delta B_t}\). 
Let \( B^{\text{min}} \) and \( B^{\text{max}} \) represent the minimum and maximum of the RB capacity, respectively. Since reverse charging to the power grid is not enabled in this work, \( M_t \) is constrained to be non-negative. Following previous constraints, range of \( \Delta B_t \) is determined as \( \Delta B_t\in [\max(- D_t, B^{\text{min}}-B_t),B^{\text{max}}-B_t ]\).

\subsection{Markov Decision Process Formulation}
\label{sec:mdp}
To further develop the system model for our study, we cast the described elements and their interactions within the framework of a time-discrete finite-horizon MDP. According to \cite{puterman1990markov}, MDP can be defined as \( (S, A, P(s_{t+1} | s_t, a_t), R, \gamma) \).
\subsubsection{State space \( S \)}The set of all possible states that the MDP can be in. In our model, the state at any time \( s_t \in S \) is represented by a tuple \( (D_t, B_t) \). EMU observes the current user load and battery capacity and makes further decisions. 
\subsubsection{Action space \( A \)}The set of all possible actions the agent can take in each state. In our work, action at any time \( a_t \in A \) is defined as the charging/discharging rate \( e_t \) of the RB: \(a_t = e_t\). Physical constraints of \( e_t \) are given by \( {e_t \in [-e^{\text{max}} ,e^{\text{max}}]}\), where \(e^{\text{max}}\) denotes the maximum charging/discharging rate of RB. Following battery operation principles, the energy exchange from the RB is determined by \({\Delta B_t = a_t   \Delta t   \eta}\), where \(\eta\) and \(\Delta t\) denote the battery charging/discharging efficiency factor and the constant sampling rate, respectively. 

\subsubsection{Transition Probability \( P(s_{t+1} \mid s_t, a_t ) \)}The likelihood of moving from the current state to the next state, given an action. In this work, we assume that the demand load \( D_t \) only depends on the user's life pattern, thus independent of the action \( a_t \) and battery level \( B_t \). The transition probability is defined as:
\begin{equation}
P(s_{t+1} \mid s_t, a_t ) = P(D_{t+1} \mid D_t ) P(B_{t+1} \mid B_t, a_t ) 
\end{equation}   
where the transition of \( B_t \) is determined by \({B_{t+1}=B_t + a_t   \Delta t   \eta}  \).
However, based on the assumption that user demand load \(D_t\) is an independent variable, we cannot directly infer the probability of state transition of  \( D_t \) over time. Thus, the overall transition probability is unknown. The model-free RL approach is appropriate for solving this problem. Such methods learn directly from experience without requiring a model of the environment.
    
\subsubsection{Reward \( R \)} The reward \({r(s_t, a_t) \in R}\) quantifies the immediate benefit of taking action \( a_t \) in state \( s_t \). In our work, the reward function is shaped as:
    \begin{multline}
    r(s_t, a_t) = \lambda  r_{\text{privacy}}(s_t, a_t) + (1 - \lambda)  r_{\text{cost}}(s_t, a_t) + \\
        r_{\text{battery}}(s_t, a_t) + r_{\text{system}}(s_t, a_t)
    \label{eq:reward_function}
    \end{multline} 
    where:
    \begin{itemize}
        \item {\( r_{\text{privacy}}(s_t, a_t) \)} measures the extent to which privacy is enhanced by EMU's control strategy. In this work, a novel reward-shaping strategy is proposed. EMU employs a sliding window (SW) to effectively detect potential load signatures from streaming load demand data. Upon the arrival of new data \( D_t\), EMU evaluates it against a dynamically computed threshold to determine if it constitutes a load signature. The decision threshold of SW, denoted by \(\tau\), for identifying a load signature is calculated at each time step \(t\) as \({\tau_t = \mu_{SW} + 3\sigma_{SW}}\), where \(\mu_{SW}\) and \(\sigma_{SW}\) represent the mean and standard deviation of the elements within the SW, respectively. The updating rules for \( SW \) upon receiving a new data point \( D_t \) are as follows:
    \begin{equation}
    SW_{\text{new}} = 
    \begin{cases} 
    \left(SW \setminus \{SW_0\}\right) \cup \{\mu_{SW}\}, & \text{if }  D_t > \tau_t, \\
    \left(SW \setminus \{SW_0\}\right) \cup \{ D_t\}, & \text{if }  D_t \leq \tau_t.
    \end{cases}
    \label{eq:sw_update}
    \end{equation}
    where \( SW_0 \) is the oldest element in \( SW \). The privacy-preserving target of our proposed load-shaping strategy is to introduce artificial load patterns and mask the original load signatures. To mitigate the impact of noise on the recognition of load patterns in data, we have introduced a signature decision threshold:
\begin{equation}
    \delta_t = \max(\delta, \tau_t)
\end{equation}
where \( \delta \) is a constant representing the noise threshold. This setup ensures that only input signals exceeding \( \delta \) are considered potential signatures. Thus, the privacy reward function is formulated by:
\begin{equation}
\begin{split}
& \qquad\qquad\qquad r_{\text{privacy}}(s_t, a_t)  =  \\
&\begin{cases}  
100+0.05( D_t-\tau_t) & \text{if }  D_t \geq \delta_t \text{ and } M_t< \delta_t, \\
-50-0.05(M_t-\tau_t) & \text{if }  D_t \geq \delta_t \text{ and } M_t \geq \delta_t,\\
20+0.05(M_t-\tau_t) & \text{if }  D_t < \delta_t \text{ and } M_t \geq \delta_t,\\
-20 & \text{if }  D_t < \delta_t \text{ and } M_t< \delta_t.
\end{cases}
\label{eq:r_privacy}
\end{split}
\end{equation}
    
    \item{\( r_{\text{cost}}(s_t, a_t) \)} reflects the savings achieved in electricity costs through strategic energy management, considering time-of-use tariffs:
    \begin{equation}
        r_{\text{cost}}(s_t, a_t) = -a_t \Delta t \eta G_t
        \label{eq:r_cost}
    \end{equation}
    where \(G_t\) denotes the electricity price at time \(t\).
    
    \item{\( r_{\text{battery}}(s_t, a_t) \)} describes the consistency in battery levels:   
    \begin{equation}
    r_{\text{battery}}(s_t, a_t) = \begin{cases} 
    \frac{10B_t}{B^{\text{max}}-B^{\text{min}}} & \text{if } t=T, \\
    0  & \text{otherwise}.
    \end{cases}
    \label{eq:r_battery}
    \end{equation}

    \item{\( r_{\text{system}}(s_t, a_t) \)} is a penalty term, designed to prevent the actions taken by the EMU agent from exceeding physical limits during the training process. Legal action ranges can be derived from physical constraints:
    \begin{equation}
        \frac{\max(- D_t, B^{\text{min}}-B_t)}{\Delta t \eta} \leq a_t \leq \frac{B^{\text{max}}-B_t}{\Delta t \eta}
    \end{equation}
    For simplicity, we use \(a^{\text{min}}_t\) and \(a^{\text{max}}_t\) to denote the lower and upper bound of \(a_t\), respectively. This penalty term is given by:  
    \begin{equation}
    r_{\text{system}}(s_t, a_t) = \begin{cases} 
    - (a_t - a^{\text{max}}_t)  & \text{if } a_t > a^{\text{max}}_t, \\
    - (a^{\text{min}}_t - a_t) & \text{if } a_t < a^{\text{min}}_t, \\
    0 & \text{otherwise}.
    \end{cases}
    \label{eq:cases_example}
    \end{equation}

    \item{\( \lambda \in [0,1]\)} represents the trade-off factor for balancing privacy and cost. By adjusting \( \lambda \), the system can prioritize either privacy or cost.
    \end{itemize}
    In order to address the reward unit inconsistency and reward coupling problem, we have implemented normalization for all reward terms except \( r_{\text{battery}} \). We utilize the normalization formula \(\hat{r}_t = \frac{r_t - \min \hat R}{\max \hat R - \min \hat R}\) to scale rewards between zero and one, where \(\hat{r}_t\) is the normalized reward term and \(\hat R\) denotes the set of all observed rewards for a particular reward term. However, \( r_{\text{battery}} \) is treated distinctly. It is awarded only at the final step of each decision cycle to emphasize maintaining the battery level consistent across decision-making periods, rather than focusing on moment-to-moment fluctuations. 

\subsubsection{Discount factor \( \gamma \)} The discount factor \( \gamma \in [0,1] \) is used to compute the expected present value of future rewards.

The overall goal within this MDP framework is to derive an optimal policy \( \pi^* \), which prescribes the best action \( a \) to take in each state \( s \) in order to maximize cumulative rewards over a decision horizon \( T \). 

\section{Methdology}
\label{sec:methodology}
\subsection{DQN-based Load-Shaping Strategy}
Our proposed load-shaping strategy PLS-DQN employs the DQN algorithm, which utilizes a model-free approach to approximate the optimal policy \( \pi^* \) in environments with unknown dynamics. DQN is particularly effective due to its dual-network architecture that helps stabilize learning and enhance the accuracy of action-value estimations.
The Q-value \( Q(s, a) \) represents the expected reward for taking an action \( a \) in a state \( s \):
\begin{equation}
    Q(s_t, a_t) = \mathbb{E} \left[ r_t + \gamma \max_{a_{t+1}} Q(s_{t+1}, a_{t+1}) \mid s_t, a_t \right],
\end{equation}

DQN algorithm uses two neural networks for the training: the Q-network and the target Q-network. The Q-network is parameterized by \( \theta \) and outputs Q-values for all possible actions in a given state.
The Q-network updates its parameters using:
\begin{multline}  
    Q(s_t, a_t; \theta) \leftarrow \\
    Q(s_t, a_t; \theta) + \alpha \left[ r_t + \gamma \max_{a_{t+1}} Q(s_{t+1}, a_{t+1}; \theta') - Q(s_t, a_t; \theta) \right]
    \label{eq:q_update}
\end{multline}
where \( \max_{a_{t+1}} Q(s_{t+1}, a_{t+1}; \theta') \) represents the maximum reward for the next state \(s_{t+1} \) as estimated by the target Q-network. The parameters of the target Q-network \( \theta' \) are updated to match the parameters of the Q-network \( \theta \) with an updating frequency \(k\).
Training of PLS-DQN involves minimizing the loss between the predicted Q-values and the target Q-values using gradient descent. The loss function is defined by:
\begin{equation}
 L = \left( r_t + \gamma \max_{a_{t+1}} Q(s_{t+1}, a_{t+1}; \theta') - Q(s_t, a_t; \theta) \right)^2
 \label{eq:q_loss}
\end{equation}
where \({r_t + \gamma \max_{a_{t+1}} Q(s_{t+1}, a_{t+1})}\) is the calculated target Q-value.
The DQN algorithm incorporates an epsilon-greedy strategy to balance the exploration of the environment with the exploitation of known values. Throughout the training, the agent chooses between selecting a random action with probability \(\epsilon\) or the action that maximizes the current Q-value with probability \(1-\epsilon\).

Training of our PLS-DQN algorithm is presented in Alg.~\ref{alg:dqn}. Initial parameters of the Q-network are randomly assigned and copied to the target Q-network (Line 1). Then the training goes into the main training loop with the initial state and SW (Lines 2-3). For each step in the episode, the PLS-DQN agent selects and executes action \(a_t\) by the epsilon-greedy strategy based on current observation \(s_t\) (Line 5). The environment receives this action, outputs the reward \(r_t\) and next state \(s_{t+1}\) (Lines 6-7). Meanwhile, SW is updated using~\eqref{eq:sw_update} (Line 8). This interaction sequence \({(s_t,a_t,r_t,s_{t+1})}\) is stored in a replay buffer (Line 9). The target Q-network uses the data sampled from the minibatch to calculate the target Q-value and the parameter of the Q-network is updated by minimizing the loss function~\eqref{eq:q_loss} using gradient descent (Lines 10-12). The parameters of the target Q-network \(\theta'\) are updated every k steps by coping values of \(\theta\) (Line 13). At last, the agent moves to the next state (Line 14).

Once the Q-network has been sufficiently trained and the Q-values are reliably approximating the true values, the optimal policy \( \pi^* \) can be directly derived from these Q-values. For each state \( s \), the optimal action \( a^* \) is determined by:
\begin{equation}
    a^* = \arg\max_a Q(s, a; \theta)
\end{equation}
where \( \arg\max_a \) selects the action \( a \) that maximizes the Q-value in state \( s \). 

\begin{algorithm}[t]
\small
\caption{Training of PLS-DQN}
\begin{algorithmic}[1]
\State Initialize \(Q(s, a; \theta\))  and \(Q(s, a; \theta')\) with \(\theta' \gets \theta\) 
\For{number of training episodes }
    \State Initialize state \(s=[D_1,B_1]\) and SW
    \For{\(t \in \mathcal{T}\)}        
        \State Select action \(a_t\) using epsilon-greedy
        \State Execute action \(a_t\), calculate reward \(r_t\) based on~\eqref{eq:reward_function}
        \State Observe \(s_{t+1}\)
        \State Updating SW based on~\eqref{eq:sw_update}
        \State Store \((s_t,a_t,r_t,s_{t+1})\) in the replay buffer
        \State Sample a minibatch from the replay buffer
        \State Calculate target Q-value using the sampled data 
        \State Update \(Q(s, a; \theta\)) based on~\eqref{eq:q_update} using gradient descent
        \State Update \(\theta' \gets \theta\) every k steps
        \State \(s_{t+1} \gets s_t\)     
    \EndFor
\EndFor
\end{algorithmic}
\label{alg:dqn}
\end{algorithm}

\begin{algorithm}[htbp]\small
\caption{Training of CNN-based Seq2Point NILM}
\begin{algorithmic}[1] 
\State \textbf{Input:} Aggregated load series data \({\{D_t \mid t \in \mathcal{T}\}}\), appliance load data \(\{{X_{i,t}} \mid t \in \mathcal{T}\}\), sequence length \(m\)
\State \textbf{Output:} Predicted appliance load data sequence \( {\{\hat{X}_{i,t} \mid t \in \mathcal{T} \}} \)
\State Initialize model parameters randomly
\State Split \(\{D_t \mid t \in \mathcal{T}\}\) into multiple overlapping windows \({\{D_t^m \mid t \in \mathcal{T}\}}\) using~\eqref{eq:sequence}
\For{number of training epochs}
    \State Feed sequence data \(D_t^m\) into the model
    \State Compute the predicted output \(\hat{X}_{i,t}\)
    \State Compute loss using~\eqref{eq:nilm_loss}
    \State Update model parameters using gradient descent
\EndFor
\end{algorithmic}
\label{alg:NILM}
\end{algorithm}

\begin{figure*}[th]
    \centering
    \subcaptionbox{}{
    \includegraphics[width = 0.45\linewidth]{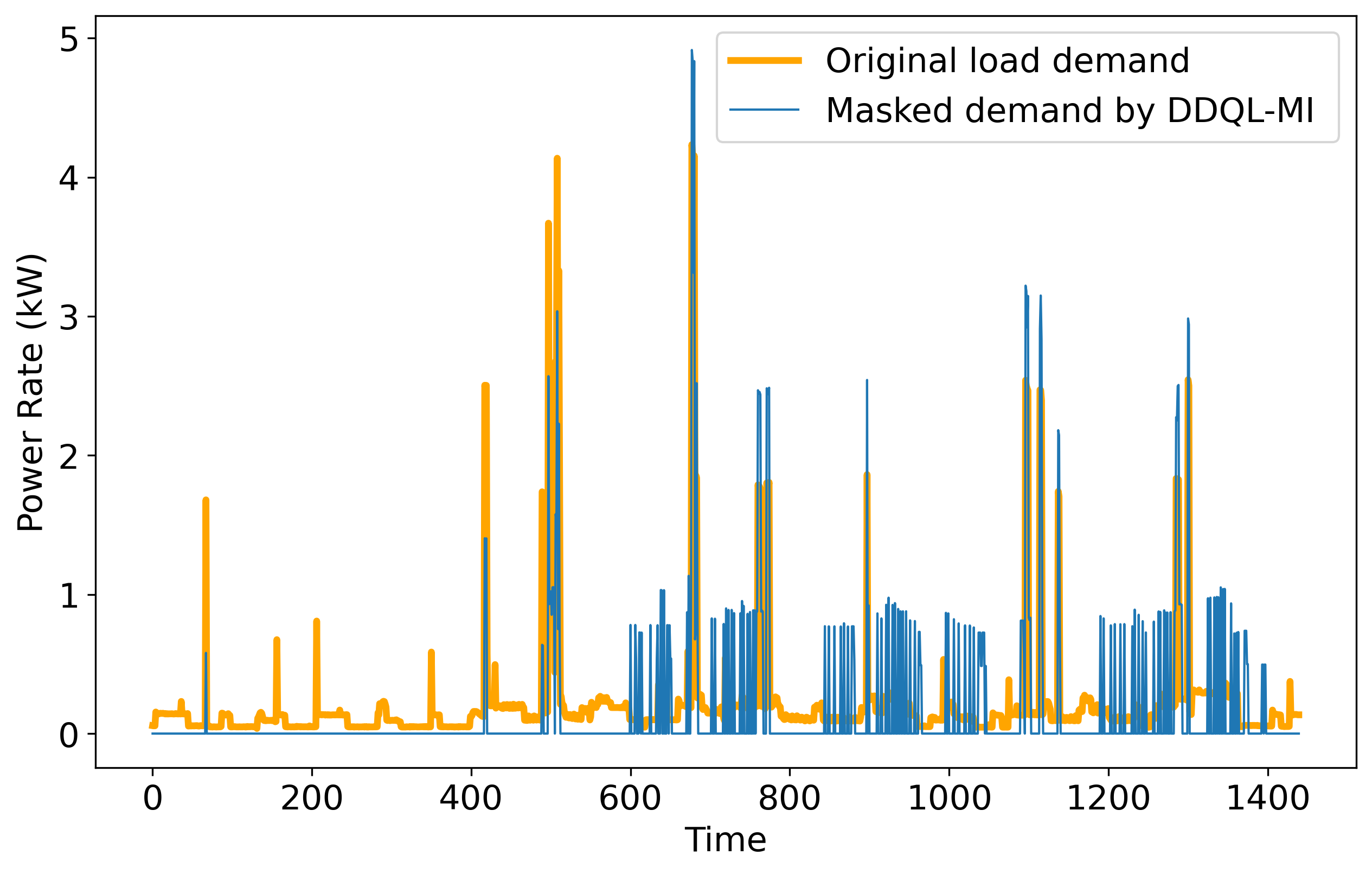}
}
    \subcaptionbox{}{
    \centering
    \includegraphics[width = 0.45\linewidth]{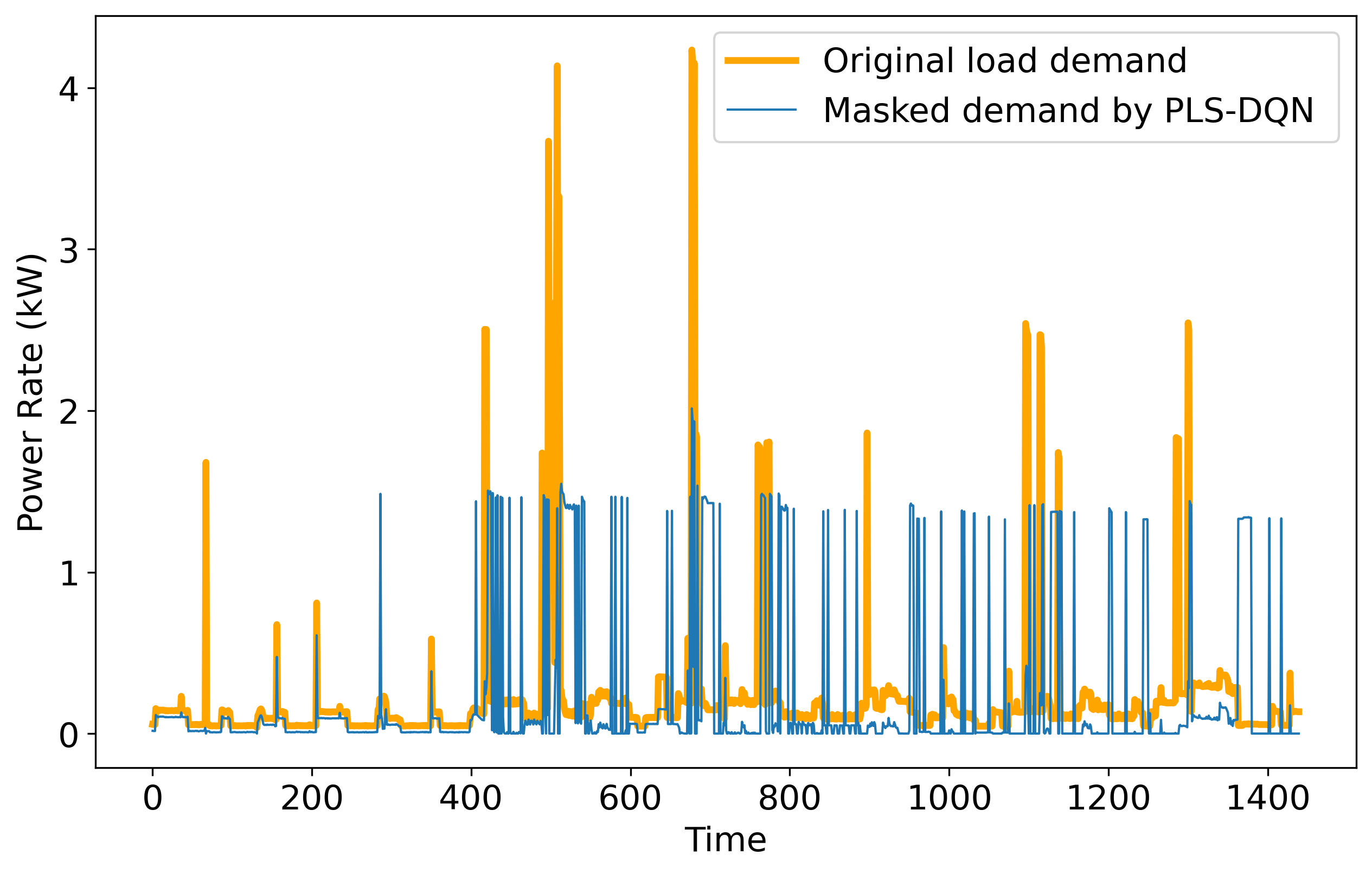}
}
    \caption{User load demand and masked load after different load-shaping strategies. (a) The result of DDQL-MI algorithm~\cite{shateri2020privacy}. DDQL-MI aims to minimize mutual information between the original load and masked load. (b) The result of our proposed PLS-DQN algorithm. PLS-DQN aims to introduce artificial load signatures while hiding original load signatures.}
    \label{fig:load_shaping}
\end{figure*}

\addtolength{\topmargin}{0.05in}
\subsection{NILM-based Adversary Model}
To evaluate the effectiveness of our proposed energy management control strategy, we incorporate a NILM system into the adversary model. This model is designed to infer the on/off status of specific appliances from the total household energy consumption data. The operation of the NILM-based adversary model can be described through the following stages:

\textit{Data Acquisition:} The adversary model receives aggregated power consumption data, expressed mathematically as:
\begin{equation}
    Y_t = \sum_{i=1}^n X_{i,t} + \beta_t
\end{equation}
where \( X_{i,t} \) denotes the power consumption of the \(i\)-th appliance at time \( t \), and \( \beta_t \) is measurement noise, assumed to be normally distributed. 

\textit{Feature Extraction:} The adversary model employs NILM techniques \( f_i(\cdot) \) to extract significant features from \( Y_t \). These features are designed to capture specific changes in power levels associated with the operational changes of the targeted appliance. In this work, we adopt a Seq2Point NILM algorithm \cite{zhang2018sequence}. This algorithm takes sequences of power data as input and outputs predictions of the power usage for a specific point in time. The sequence data at time \(t\) is represented by the nearest \(m \) elements: 
\begin{equation}
    D_t^m=\{D_k \mid k = t - \frac{m-1}{2}, \ldots, t + \frac{m-1}{2}\}
    \label{eq:sequence}
\end{equation}
For simplicity, \(m \) is assigned an odd value. The underlying framework of the NILM is based on convolutional neural networks~\cite{albawi2017understanding} (CNN), which effectively capture spatial hierarchies in data for feature extraction and subsequent prediction.

\textit{Objective Function}: The objective of the NILM is to maximize the accuracy of its predictions. To achieve this, NILM focuses on minimizing the mean square error (MSE) loss by gradient descent. The loss function is defined as:
\begin{equation}
     L(\hat{X}_{i,t}, X_{i}) = \sum_{t=1}^{T}(\hat{X}_{i,t}-X_{i,t})^2
\label{eq:nilm_loss}
\end{equation}
where \( \hat{X}_{i,t} \) represents the prediction of the \(i\)-th appliance usage at time \( t \) by the NILM.

Training of the CNN-based Seq2Point NILM algorithm is presented in Alg.~\ref{alg:NILM}. 
This NILM algorithm takes aggregated load data \(\{D_t \mid t \in \mathcal{T}\}\), appliance load data \(\{{X_{i,t}} \mid t \in \mathcal{T}\}\) and sequence length \(m\) as input (Line 1). The output is the predicted appliance consumption data sequence \( \{\hat{X}_{i,t} \mid t \in \mathcal{T} \} \) (Line 2). The parameters of the network are assigned random numbers (Line 3). The input series are split into sequences \(D_t^m\) with the length of \(m\) using~\eqref{eq:sequence} (Line 4). For every training epoch, the model takes sequence data \(D_t^m\) as input and outputs the prediction result \(\hat{X}_{i,t}\) (Lines 5-7). Then, the loss is calculated through~\eqref{eq:nilm_loss} and the parameters of the network are updated by gradient descent (Lines 8-9). 
    
\textit{Appliance State Detection}: The adversary model uses a classification function \( u_i(\cdot) \) to determine the on/off state of the appliance based on the extracted features:
\begin{equation}
    \hat{C}_{i,t} = u_i(f_i(Y_t))
    \label{eq:classfication}
\end{equation}
where \( \hat{C}_{i,t} \) indicates the predicted state of the \(i\)-th appliance at time \( t \).

\section{Results and discussion}
\label{sec:results}

\subsection{Dataset and Parameter Assignments}
In this work, the UK recording domestic appliance-level electricity (UK-DALE) dataset \cite{kelly2015uk} is used to train and test the models. UK-DALE was built at a sample rate of 1/6 Hz for individual appliances. We employ sample rate \(\Delta t = 1 \text{min} \) primarily to reduce computational complexity, while simultaneously considering the frequency of control actions. The length of an episode is defined as the total number of minutes in a day: \(T=1440\). The noise threshold \(\delta\) is 0.5 kW. The assignment of battery parameters is as follows: \(B^{\text{min}}=0\), \(B^{\text{max}}=1.5\) kWh, \(e^{\text{max}}=5\) kW, \(\eta=1\), \(B_1=B^{\text{max}}\). We consider Economy 7 electricity tariffs in the UK: \pounds 0.304 kWh from 7:00 to 24:00 and \pounds 0.132 kWh from 0:00 to 7:00. Hyper-parameters of PLS-DQN are as follows: length of SW is 5, learning rate \(\alpha\) is \(10^{-4}\), buffer size is \(10^{6}\), the batch size is 32, initial exploration rate \(\epsilon_\text{initial}\) is 1.0 and final exploration rate \(\epsilon_\text{final}\) is 0.05, training episodes is 1500, steps in a single episode is 1440, discount factor \(\gamma\) is 0.99, target network update frequency \(k\) is \(10^{4}\), the algorithm's architecture is a multi-layer perception (MLP) with 2 hidden layers, each hidden layer containing 64 neurons, with rectified linear unit (ReLU) activation functions applied to each layer. Appliances and aggregated consumption data from 0:00 to 23:59 on May 29th, 2013, in House 1 of the UK-DALE dataset are utilized to establish the DRL environment.

\begin{table*}[t]
    \caption{Numerical Results}
    \centering
    \vspace{-2mm}
    \begin{tabular}{c|ccc|ccc|ccc}
        \toprule
        \multirow{2}{*}{NILM on} & \multicolumn{3}{c|}{Kettle} & \multicolumn{3}{c|}{Toaster} &\multirow{2}{*}{Cost} &\multirow{2}{*}{Remain battery}&\multirow{2}{*}{Battery-compensated}\\
        \cmidrule{2-7}
         & Precision & Recall & $F_1$   & Precision &Recall & $F_1$ & (\pounds) &level& cost (\pounds)\\
        \midrule{}
        Original load demand    & 0.500   & 0.895   & 0.642       & 0.286   & 0.250   & 0.267 & | &| & |\\
        DDQL-MI with $\lambda=1$     & 0.235   & 0.842   & 0.368       & 0.123    & 0.500   & 0.198 &-0.324& 1.94\%  & 0.123 \\
        PLS-DQN with $\lambda=1$     & \textbf{0.020}   & \textbf{0.053}   & \textbf{0.029}      & \textbf{0.031}   & \textbf{0.062}    & \textbf{0.042} & -0.085 & \textbf{65.7\%} & \textbf{0.071}\\
        \midrule{}
        PLS-DQN with $\lambda=0.7$     & 0.042   & 0.053  & 0.047       & 0.040    & 0.062   & 0.049 & -0.003 & 92.7\% & 0.030 \\
        PLS-DQN with $\lambda=0.3$       & 0.043   & 0.053   & 0.048       & 0.056    & 0.062   & 0.059& 0.025& 100\% & 0.025\\
        PLS-DQN with $\lambda=0$         & 0.481  & 0.684   & 0.565       & 0.250    & 0.125   & 0.167& 0.023 & 100\% &0.023\\
        \bottomrule
    \end{tabular}
    \label{tab:classfication}
\end{table*}

\begin{figure}[tbp]
    \centering
    \includegraphics[width = 0.9\linewidth]{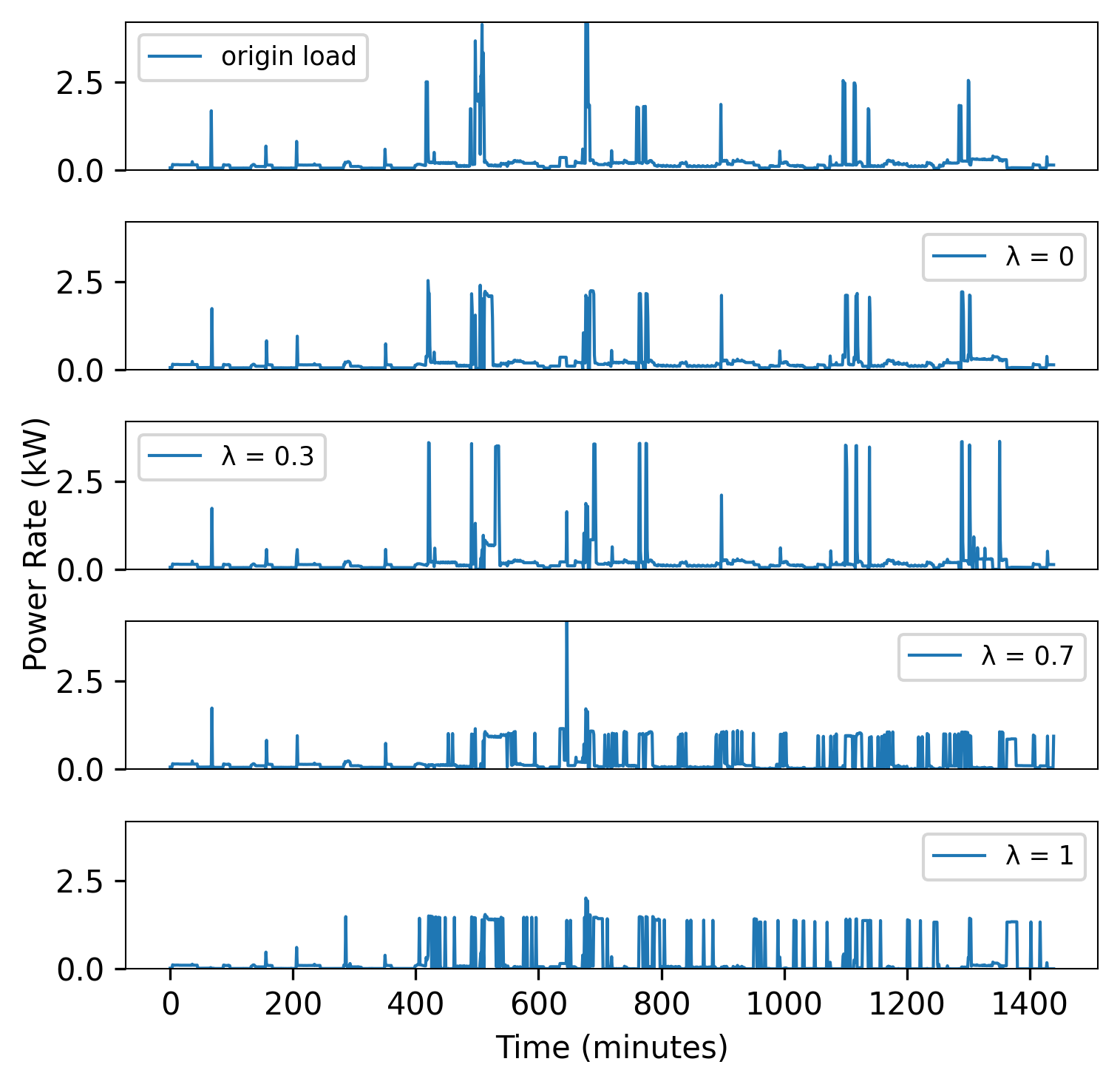}  
    \caption{The impact of the privacy-cost trade-off factor \(\lambda\) on results of PLS-DQN algorithm.}
    \label{fig:dqn_lambda}
\end{figure}
The Seq2Point NILM algorithm employs a CNN architecture that consists of two convolutional layers and a fully connected layer. The first convolutional layer features 16 output channels with a kernel size of 5, stride of 1, and padding of 2. The second convolutional layer increases the output channels to 32, maintaining the same kernel size, stride, and padding. Both layers are followed by ReLU activation functions and a global average pooling layer precedes the final output layer. The network culminates in a fully connected layer with a single output unit. Optimization of the network is facilitated by the Adam optimizer with an initial learning rate of 0.005 and decreases to 0.001 during the training process. The number of training epochs is \(10^5\). The sequence length is 5. We utilize data on appliances and aggregated consumption recorded from 00:00 on May 19th, 2013, to 23:59 on May 28th, in House 1 of the UK-DALE dataset to train our NILM algorithm. Specifically, we focus on predicting the usage patterns of kettles and toasters as they are often related to users' daily activities.

\subsection{Performance of PLS-DQN}
Figure~\ref{fig:load_shaping} shows the shaped load pattern under our proposed PLS-DQN algorithm and the state-of-the-art DDQL-MI algorithm~\cite{shateri2020privacy}. We observe that almost all the power peaks are hidden under the PLS-DQN strategy. Meanwhile, artificial load signatures above 1 kW are introduced. 

Figure~\ref{fig:dqn_lambda} illustrates how the trade-off factor \(\lambda\) controls our load-shaping strategy. With \(\lambda\) close to zero, the algorithm tends to avoid doing battery charging actions to reduce cost. In this situation, utility cost is maximally reserved at the cost of potential privacy risk. With \(\lambda\) becoming closer to one, the major peaks of the household are gradually masked and other peaks appear.

\begin{figure}[t]
    \centering
    \includegraphics[width = 0.9\linewidth]{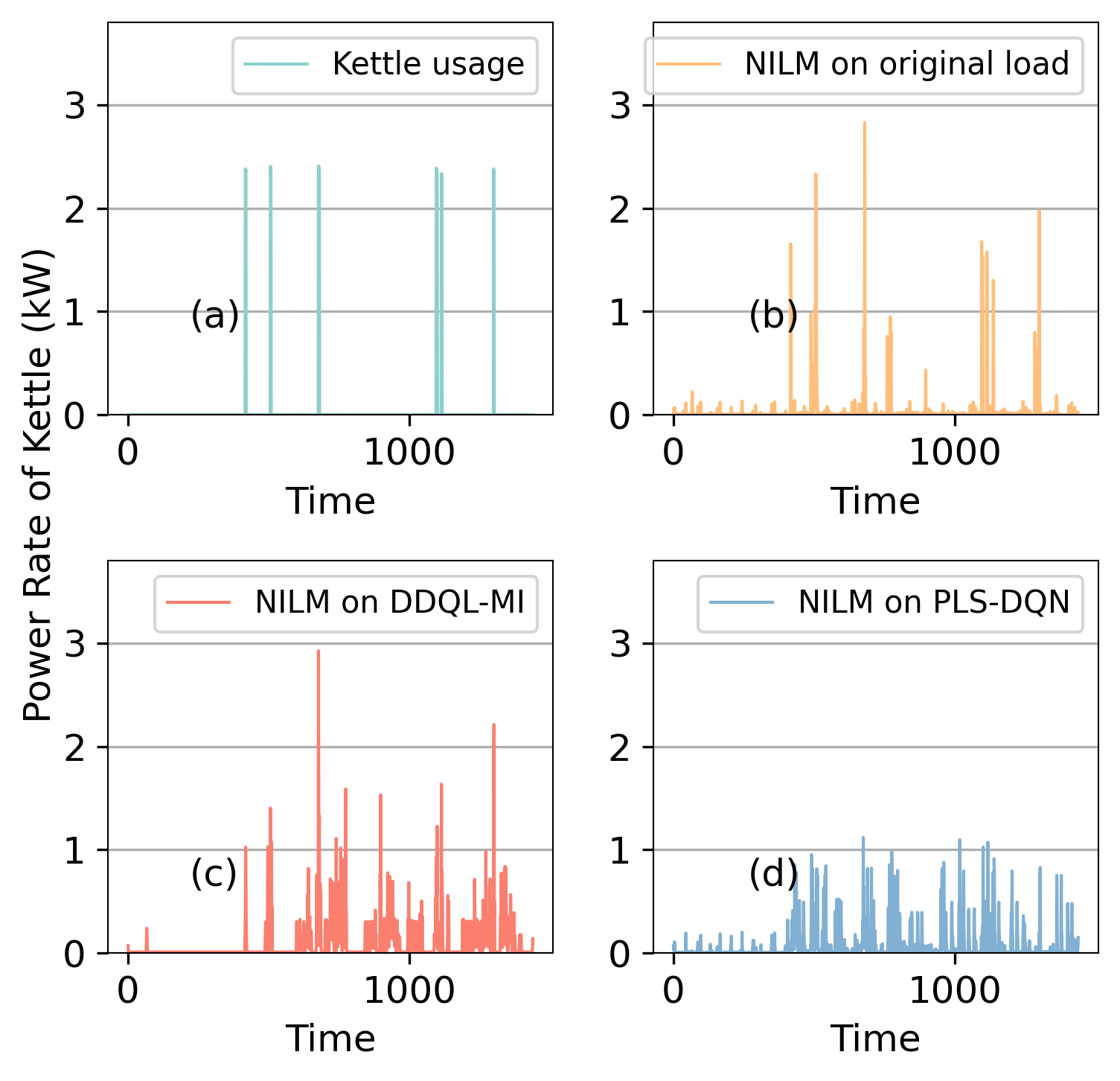}
    \caption{Load disaggregation of kettle.}
    \label{fig:kettle}
\end{figure}

\begin{figure}[tbh]
    \centering
    \includegraphics[width = 0.9\linewidth]{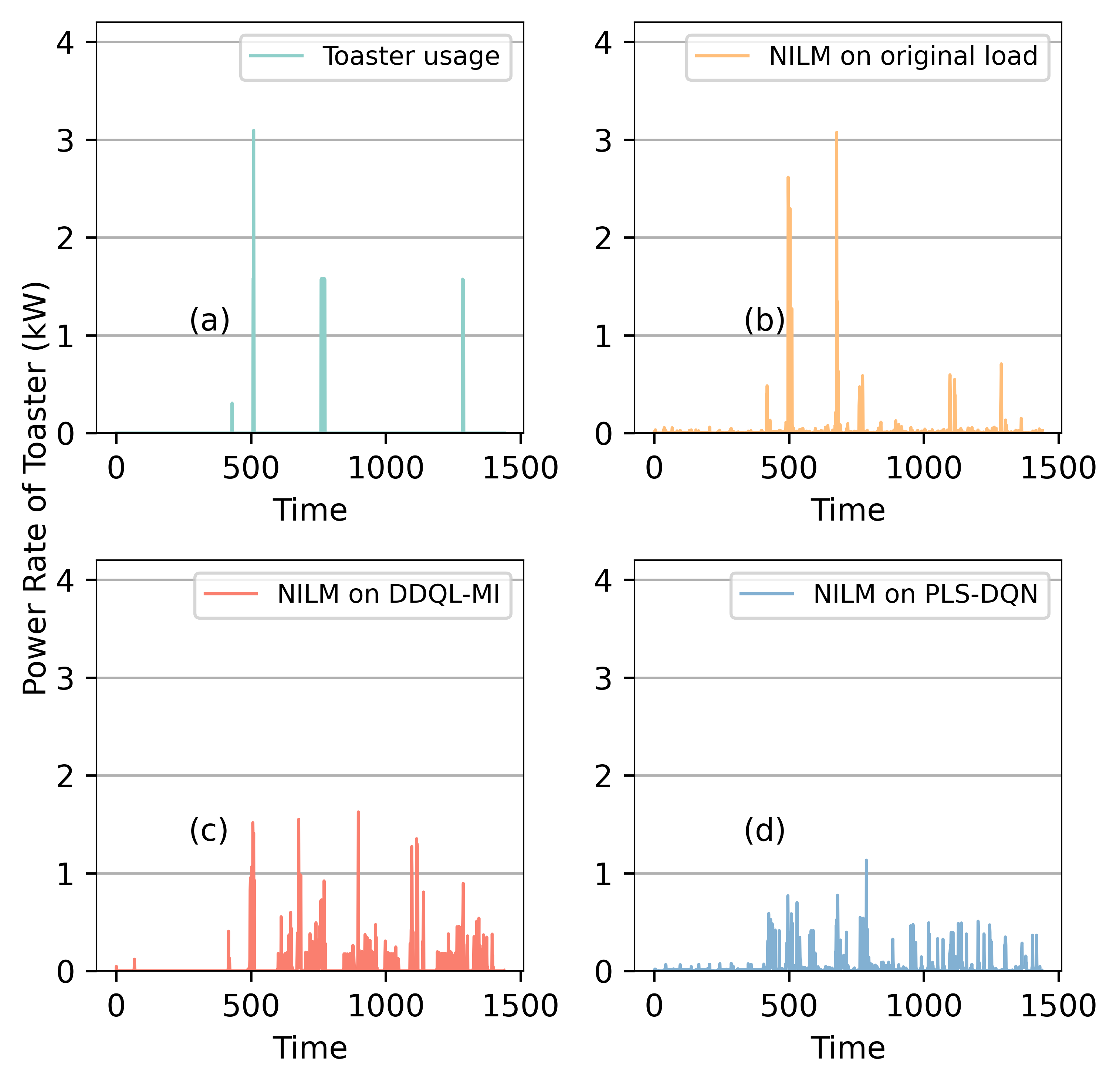}
    \caption{Load disaggregation of toaster.}
    \label{fig:toaster}
\end{figure}

To explore the performance of the PLS-DQN strategy in a real-world scenario, we trained and utilized the NILM algorithm as an adversary. We input the original user load demands, the load demands processed by the DDQL-MI algorithm and the PLS-DQN algorithm into NILM. As illustrated in Fig.~\ref{fig:kettle}, the trained NILM was nearly able to capture the operational times of the kettle, although it also produced some false positives. For the load curves protected by the DDQL-MI algorithm, NILM could still detect the kettle's operation, but with more false positives compared to the original curve. In Fig.~\ref{fig:kettle}-(d), for the load curves protected by the PLS-DQN algorithm, NILM was almost unable to correctly identify the actual operation times of the kettle. As shown in Fig.~\ref{fig:toaster}, facing appliances like toasters with multiple operation settings, the NILM algorithm struggled to accurately determine their operating power levels. However, it could still effectively capture the start times of appliances from unprotected original electrical load curves in Fig.~\ref{fig:kettle}-(b). 

The numerical results of the PLS-DQN algorithm are measured by the adversary NILM model. We apply Precision, Recall, and $F_1$ score as the evaluation metrics. If the NILM can easily classify the on/off state from the masked load data, the load-shaping algorithm is considered vulnerable. A lower score of the NILM represents a better performance of the load-shaping algorithm. We established a straightforward rule for the classification function \(u_i(\cdot)\): If the input exceeds 0.5 kW, it is classified as `on'; otherwise, it is classified as `off'. As shown in Table~\ref{tab:classfication}, PLS-DQN proves to be effective on both tasks. Also, PLS-DQN has the ability to maintain the battery consistently in the end of the decision cycle. As different algorithm ends up with different battery level, the cost is not comparable. We calculated battery-compensated cost by adding the cost of charging the battery to the maximum for all algorithms and settings. The conclusion is that PLS-DQN outperforms the state-of-the-art DDQL-MI algorithm against a NILM-based classification adversary model both in privacy protection and in electricity cost.

\section{Conclusions}
\label{sec:conclusion}
In this work, we proposed a deep reinforcement learning-based load-shaping algorithm named PLS-DQN. It utilizes a rechargeable battery to mask the household electricity load. PLS-DQN outperforms the state-of-the-art algorithm against a NILM-based classification algorithm. The results demonstrate that the PLS-DQN algorithm not only enhances user privacy but also effectively preserves battery consistency after the operating cycle, achieving these outcomes with reduced cost.

\bibliographystyle{IEEEtran}
\bibliography{ref}

\end{document}